astro-ph/9712251 18 Dec 1997

# What can *BeppoSAX* do about the 2-10 keV cosmic background ? A progress report.

L.Chiappetti[a],G.Cusumano[b],S.Del Sordo[b], M.C.Maccarone[b], T.Mineo[b] and S.Molendi[a]

[a]Istituto di Fisica Cosmica e Tecnologie Relative (IFCTR/CNR)
via Bassini 15, I-20133 Milano, Italy
[b]Istituto di Fisica Cosmica e Applicazioni di Informatica (IFCAI/CNR)
via La Malfa 153, I-90146 Palermo, Italy

We report the current status of the analysis of the MECS background using the entire dataset of the *BeppoSAX* Science performance Verification Phase. We have collected 360 ks of dark Earth instrumental background, 470 ks of bright Earth background and 1100 ks of blank field data. We are attempting to model the instrumental background in terms of its various components (in particular the spatial modulation of the residual contamination by the built-in Fe calibration sources), and then use this model, and the information on the vignetting and the PSF to derive the cosmic background in the 2-10 keV range.

## 1. INTRODUCTION

The origin of the diffuse hard X-ray background is commonly interpreted as integrated emission of unresolved X-ray sources (possibly AGNs) along the line of sight (for a review see [1] and refs. therein).

The MECS [2] instrument on board *BeppoSAX* [3] is particularly suited for the study of the cosmic X-ray background in the 2-10 keV range thanks to its low and stable instrumental background.

*BeppoSAX* was launched on 30 April 1996 on a low Earth equatorial orbit. After the spacecraft and payload commissioning phase, the Science Verification Phase (SVP) began on 12 July 1996. During this phase a number of science and calibration pointings were made either with the NFIs or the WFCs being prime. The data rights belongs to the *BeppoSAX* hardware teams for one year. Most SVP pointings were executed during July and August 1996, then they were interleaved with the Core and Guest Observer Programmes and the last residual SVP pointings were executed on 4 December 1996.

## 2. DATA REDUCTION

The SVP data amount to 210 Observing Periods (OPs). The MECS instrument (all three units) was on for 4.2 Ms out of 5.8 Ms with data present.

The work described here is part of a systematic analysis of the verification of the behaviour of the instrument, which involved among others the following steps :

- preliminary data cleaning (elimination of spurious telemetry packets)
- adding fake (nominal) attitude files when missing
- generation of standard time windows, in particular of the intervals when the target was not occulted by the Earth, and when the instruments were pointing at the dark and bright Earth respectively (XAS program `saxauxcalc` [4])
- generation of XAS standard gain histories for the correction of time dependent gain variations

The preliminary data reduction for the study of the background was then conducted as follows. We divided the useful field of view in 35 boxes (each one being 31×31 0.15 mm pixels, i.e. 4.8 arcmin side) as shown in Fig. 1. The box size was a a compromise between a reasonable statistics, and a fair spatial resolution.

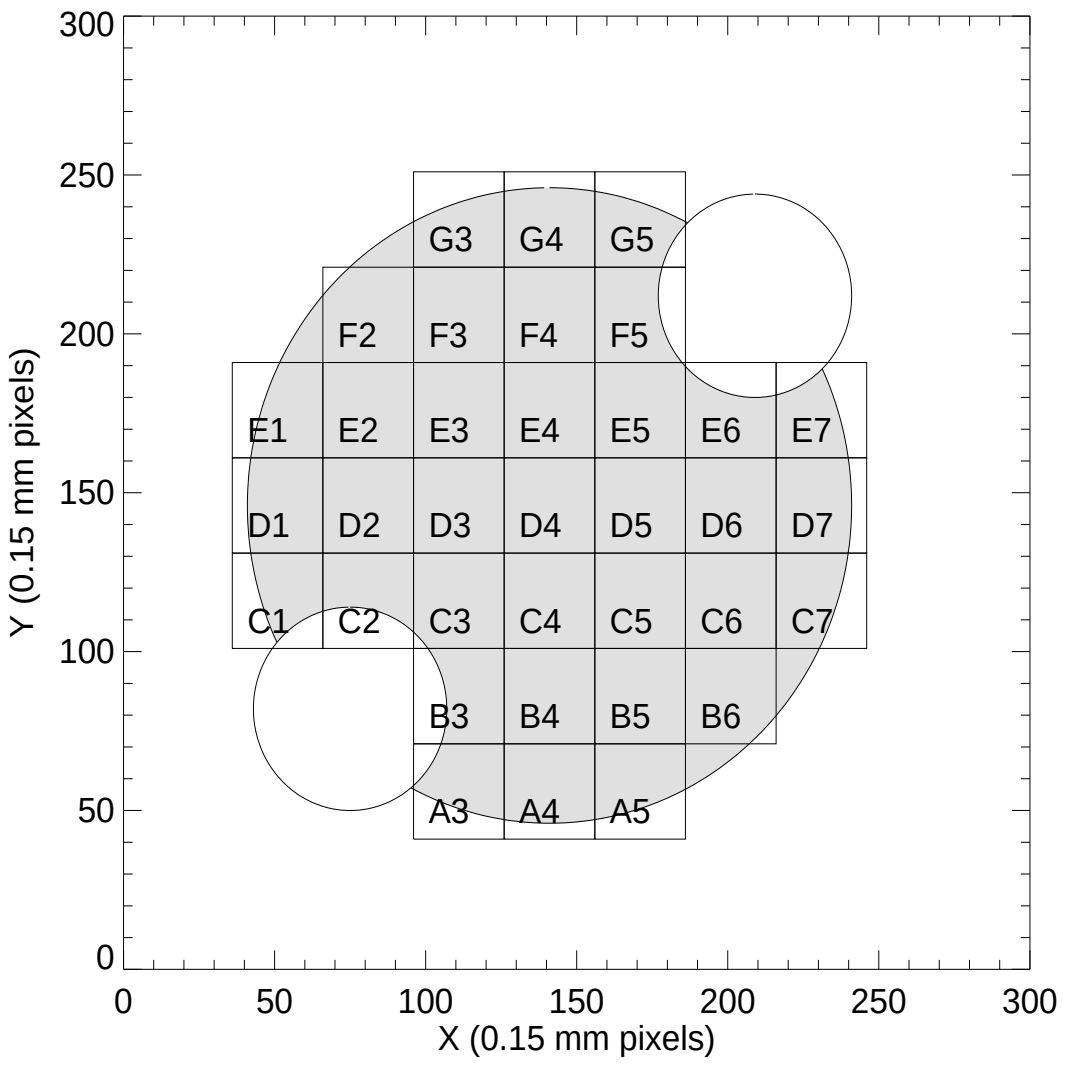

Figure 1. The boxes used for spectra integration overlaid over the MECS entrance Be window. The small circles indicate the avoidance regions around the built-in $^{55}Fe$ sources. Data are accumulated only inside the shaded area.

For each OP, for each of the 3 MECS units and for each of the 3 kind of time windows (unocculted source, dark Earth or bright Earth) we accumulated an energy spectrum in each box (for a total of more than 60,000 spectra).

We looked for any anomaly in the automatic procedure and rejected the relevant spectra. For unocculted source data we also did a coarse selection of blank fields. We rejected immediately all OPs containing a bright target (i.e. visible at ratemeter level), then visually inspected images of the remaining OPs and rejected those showing serendipitous sources. The final exposure times for each "class" (blank,dark,bright) are listed in Table 1 (values are approximated, there are little differences by MECS unit)

Table 1
Breakdown of data by time window

| Condition | exposure time (ks) |
|---|---|
| target not occulted | 6688[a] |
| of which blank fields | 1100 |
| dark Earth | 360 |
| bright Earth | 470 |

[a] this value is not an exposure time, but the sum of the duration of the nominal time windows

We then added all spectra of the same box, same MECS unit and same class, obtaining 9 families of 35 spectra each, with the exposure times given in Table 1. At this point we did either :

combine the 35 spectra of the same class and MECS unit, to obtain spectra averaged over the entire field of view (see Fig. 2 and Fig. 4), or compute the count rate in a given PHA band from each spectrum (obtaining 9 sets of 35 numbers) and interpolate such count rates as a function of position, obtaining a map of the background (see Fig. 3).

## 3. INSTRUMENTAL BACKGROUND

### 3.1. Phenomenological description

The instrumental background (see Fig. 2) appears to be constituted by a number of components :

(a) the noise of the PhotoMultiplier Tube (PMT) , different for each MECS unit but always below PHA channel 25 ; being outside the useful energy range, it will not be considered further.

(b) a continuum appearing as a gently sloping plateau below 4 keV, and another plateau at an higher level above at least 4.5 keV. The continuum is terminated by the response high energy cutoff whose precise position (> 10.5 keV) depends on the gain setting of the individual MECS unit. This component is not position dependent.

(c) an instrumental feature appearing as a step, or more likely a small line between 4 and 4.5 keV

(d) a line feature peaked around the calibration source ($^{55}Fe$) energy (5.9 keV), definitely broader than energy resolution (or multi-peaked) and spatially modulated (see Fig. 3). There is perhaps an hint of a minor feature at 6.4 keV.

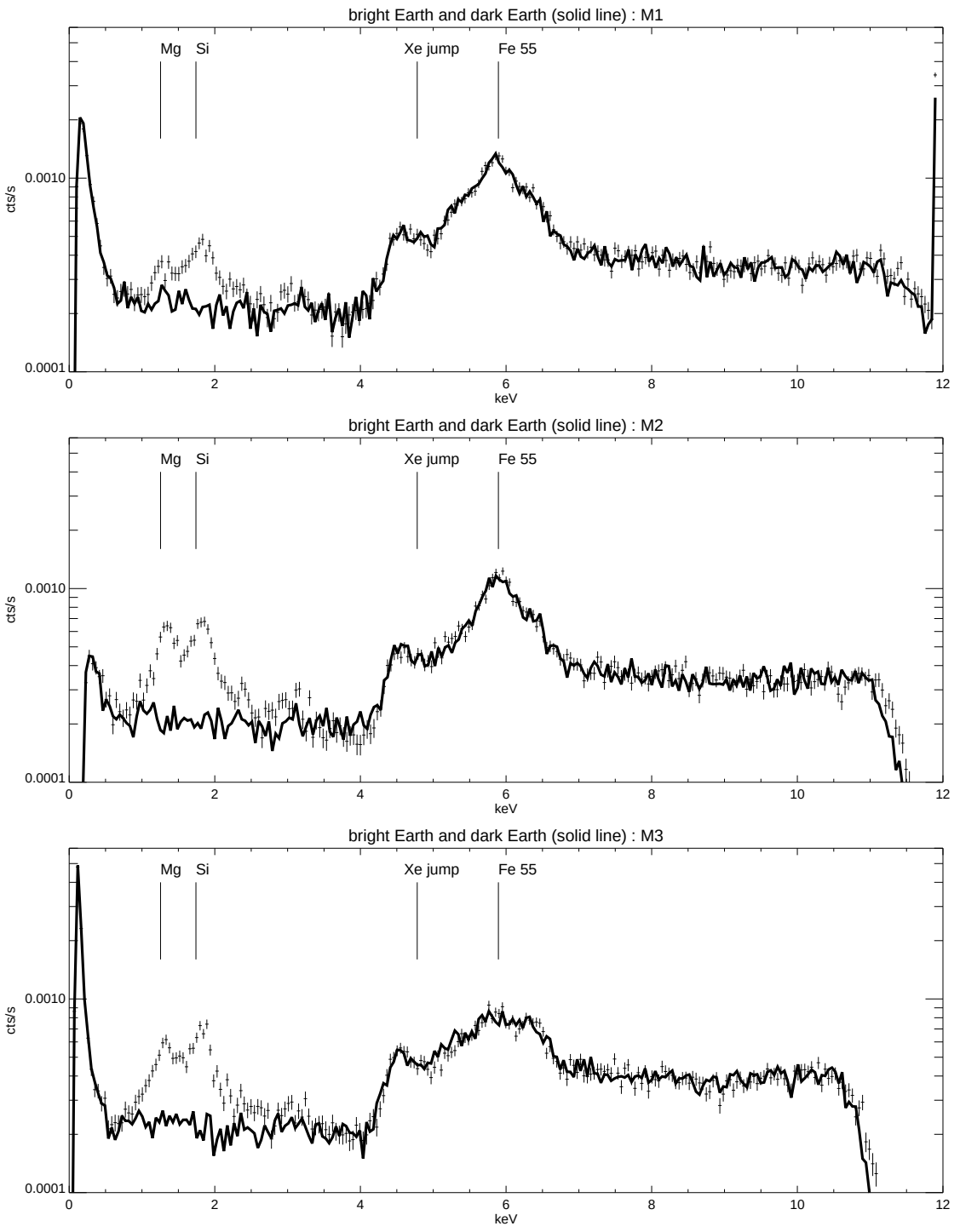

Figure 2. The spectrum of the dark Earth background (solid line) and of the bright Earth background (error bars) for the 3 MECS units, referred to the entire FOV.

(e) some weak line features below 4 keV, present only in bright Earth spectra and spatially modulated according to the vignetting by the optics. The Mg and Si lines show up clearly. They can be attributed to Sun albedo or to atmospheric fluorescence in the case of Ar, and will be not considered further. Note that, with the exception of these features, the dark and bright Earth spectra are otherwise identical.

(f) whatever remains (see Fig. 4) is the cosmic background, which is obviously vignetted by the optics.

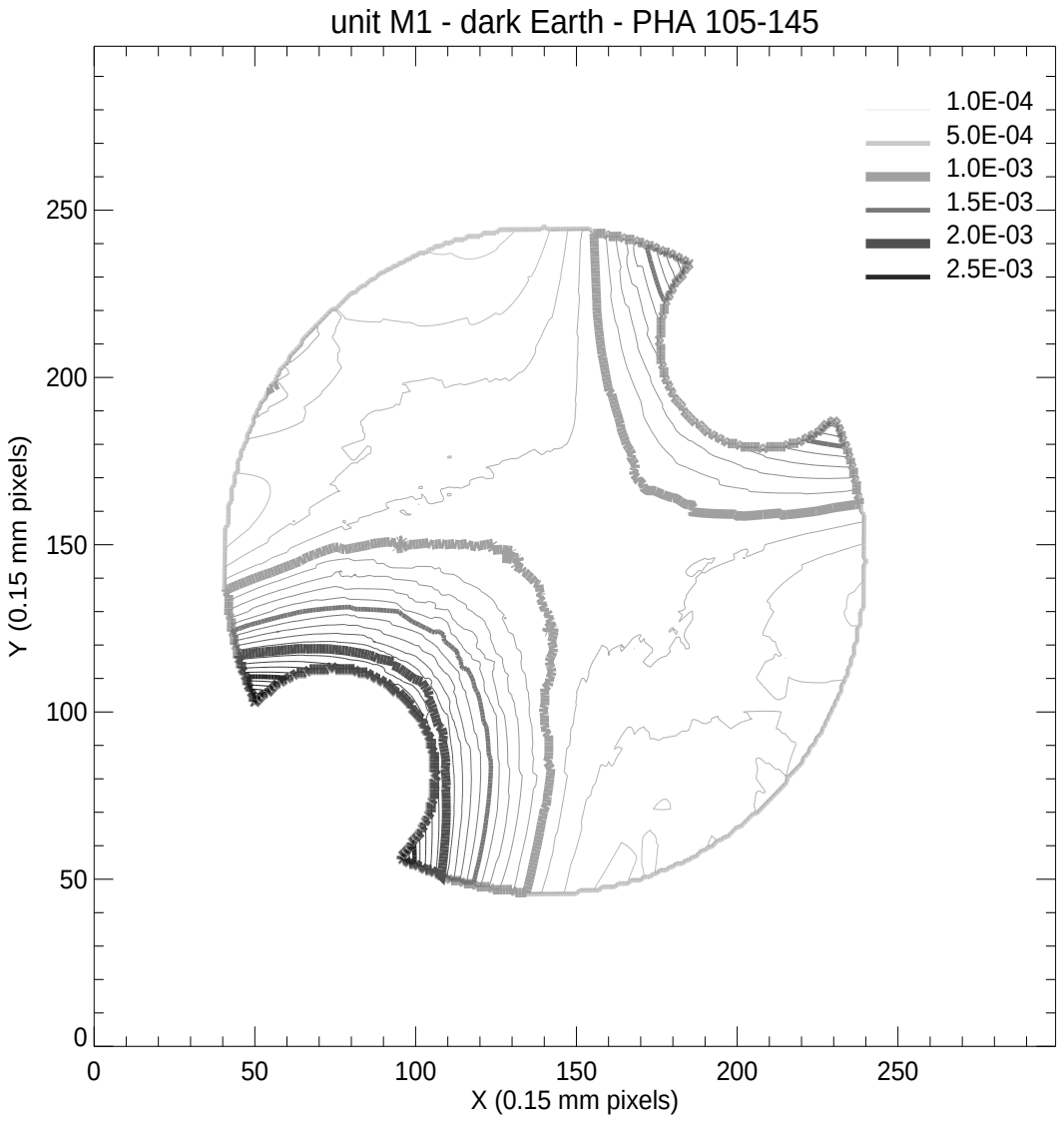

Figure 3. Example of spatial modulation of the instrumental background : count rate in the Fe-band (4.8-6.7 keV) of dark Earth data for detector M1.

Component (d), the residual Fe feature, deserves some further description. It is spatially modulated (see Fig. 3) in the form of an intensity decrease as a function of distance from the calibration source positions. An inspection of the individual "box" spectra shows that the line is narrow and compatible with the energy resolution for positions close to the calibration source, where it is also more intense. Note that the intensity of this feature (maximum gross rate for outer boxes 0.52-$1.07\times10^{-3}$ cts/s according to MECS source and unit) although clearly visible, is a tiny fraction (~0.1%) of the calibration source intensity.

The full FOV background rate in the 2-10 keV band is respectively 0.10 and 0.066 cts/s for blank field and dark Earth spectra.

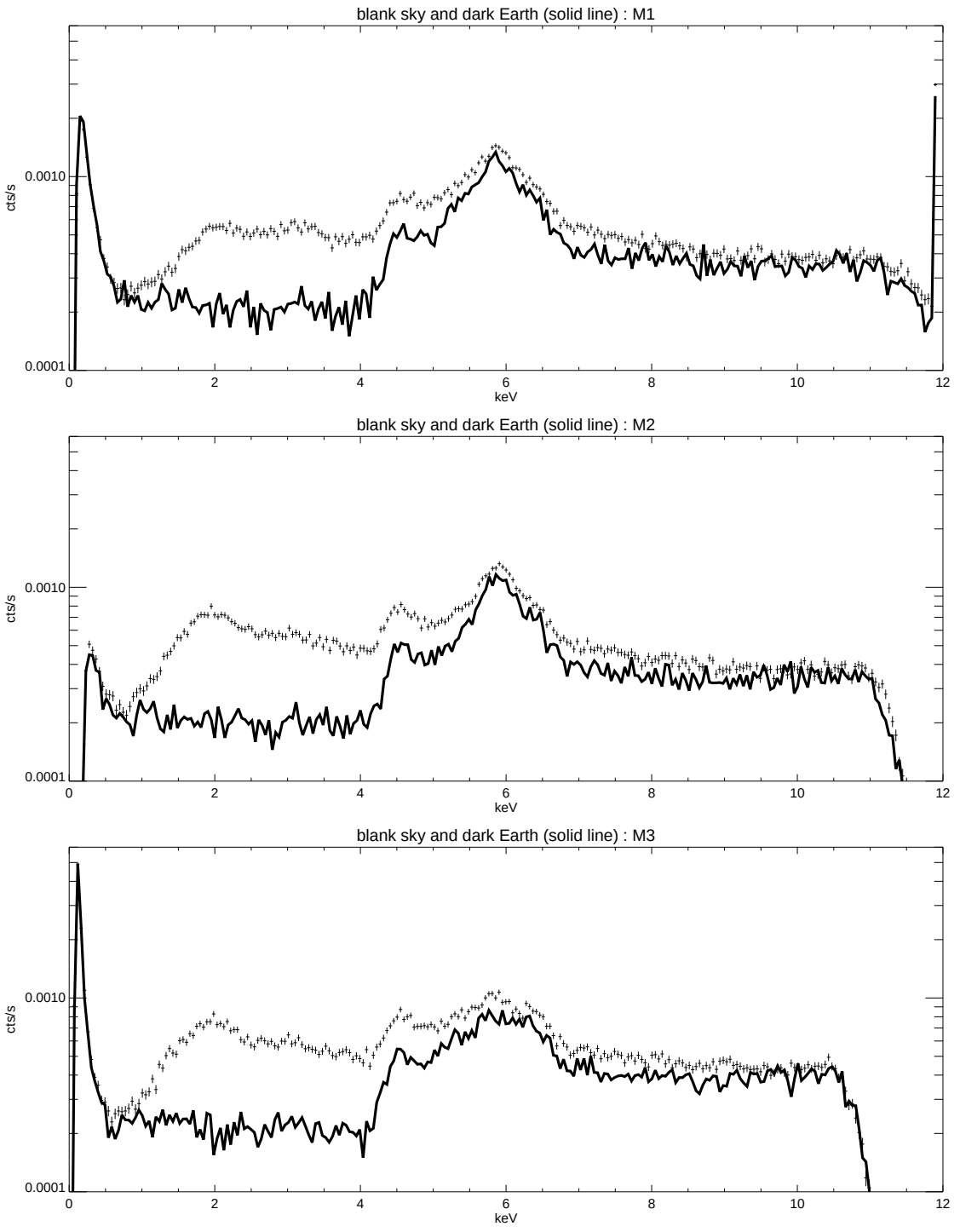

Figure 4. The spectrum of the dark Earth background (solid line) and of the blank field background (error bars) for the 3 MECS units, referred to the entire FOV.

### 3.2. The model

Ideally one would like to develop a complete model of the instrumental components of the background. This could be subtracted from the blank field spectra (see Fig. 4) to derive the net spectrum of the cosmic background, as well as applied for background subtraction of celestial sources.

We originally attempted to develop an empirical model in the following form. We first modelled components (b) and (c) as a continuum fitting two linear stretches of the form a+bE (one below 4 keV, and the other one from 4.5 to 5 keV and from 7 to 10.5 keV) joined by a linear interpolation (4-4.5 keV). This component is independent on position.

We subtracted the model from the spectra of the individual boxes, and generated maps of intensity in the Fe band (component (d)). We found out that the spatial modulation is well represented by the form $k_1/r_1^2 + k_2/r_2^2$ where $r_i$ are the distances from the individual calibration source positions, and the normalizations $k_i$ scale as the relative intensities of the calibration sources. We have not yet taken into account the decay of the $^{55}$Fe sources with time.

While we can model the intensity of the line as a function of position, we however encountered difficulties in modelling its shape. We tried a simple Gaussian shape with width $\sigma$ constant or function of intensity (line broader when weaker). We achieve a qualitative agreement, but if we subtract our model of components (b)+(c)+(d) from real data we have unsatisfactory residuals.

We tried another approach, which involves tentatively some physical explanation of components (c) and (d). We first remind of the principles of operation of a GSPC (see Fig. 5).

X-ray photons interact with the Xenon atoms in the gas cell, and give rise by photoelectric effect to an electron cloud which drifts, under the influence of the applied electric field, toward the scintillation region, where the cloud is converted into UV photons, finally collected by the PMT. X-rays above 4.78 keV may give rise to a fluorescence photon, which can interact separately or escape. According to the direction taken by the fluorescence photon, the electron clouds of the primary interaction and of the fluorescence photon may arrive at different times in the scintillation region, and give rise to bursts of light of different duration (Burst Length, BL), which will be detected anyhow as a single event (except those with longer BL which are usually rejected by software).

The above ordinary effects are duly taken into account in the instrument response matrix. We believe that components (c) and (d) can be explained by some marginal effects.

If the primary charge deposit is absorbed by attachment to the entrance window (which may happen for interactions close to it), one is left only with fluorescence photons. Considering the fine structure of the Xe L edge (which gives four fluorescence components, of which the two most intense are at 4.1 and 4.4 keV), component (c) must be interpreted as a blended line feature.

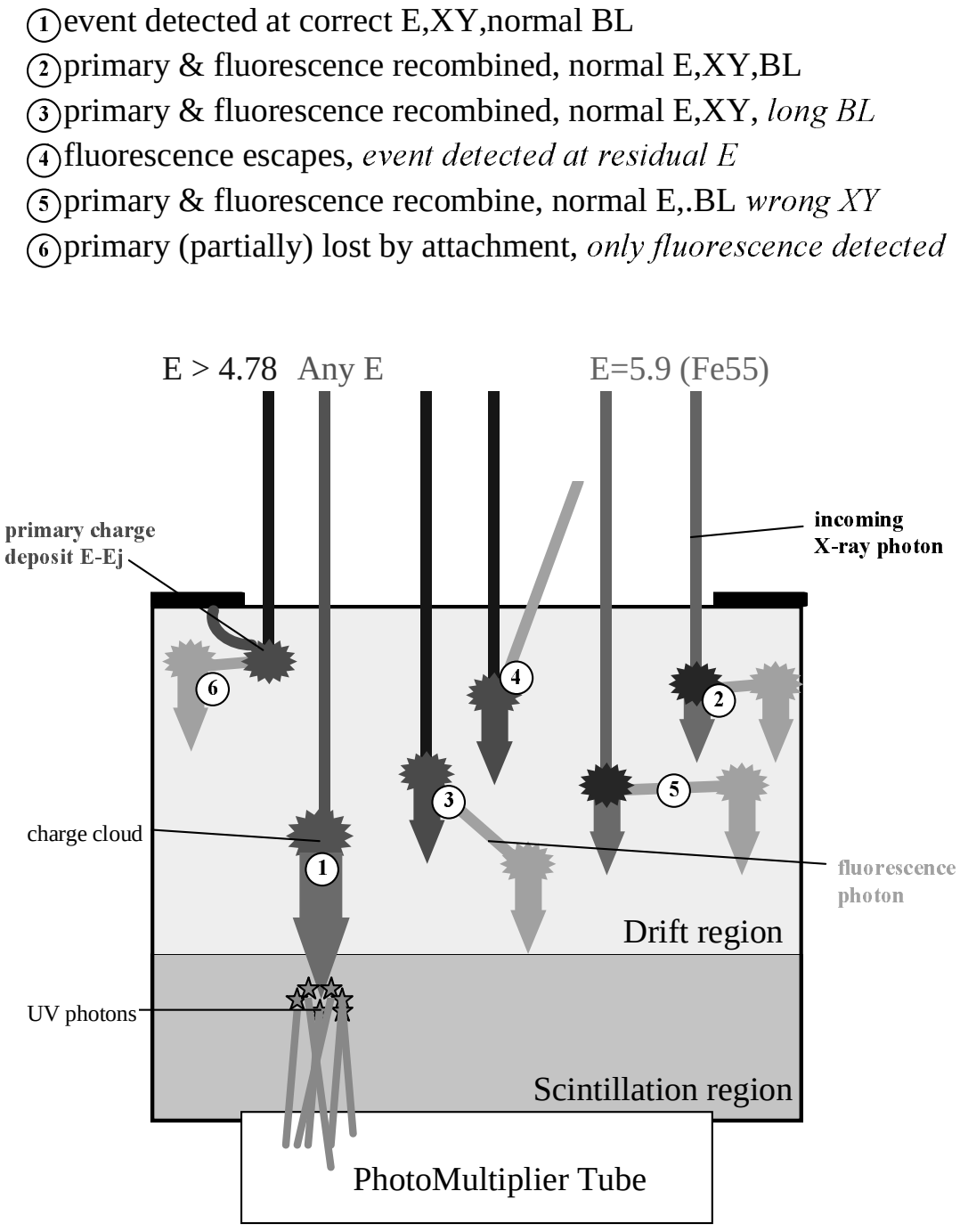

Figure 5 : Diagram representative of the main kind of interaction of an X-ray photon within a GSPC.

For what concerns component (d) one shall take into account that $^{55}$Fe 5.9 keV photons are above the Xe L edge. There is a non negligible probability that the fluorescence photon travels a long distance in horizontal direction, resulting in the detection of an event with nearly correct PHA and BL, but a displaced position (X,Y instead of the calibration source position $X_C,Y_C$). One shall furthermore consider that standard data reduction corrects each event PHA for gain disuniformity, normalizing it to the detector centre [2, fig. 3]. Therefore misplaced photons are assigned the "wrong" energy.

We attempted a modelling using as input two spatially dependent Gaussian lines with centroid at 5.9 keV, normalization proportional to $k_i$ and width σ given by the nominal resolution; we then compensated for the wrong gain correction using a factor $g(X_C,Y_C)/g(X,Y)$. This is able to reconstruct qualitatively the shapes of the observed features, but still underestimates the total width. Some further simulation work is therefore necessary.

Concerning component (b), the instrumental continuum, we have suspended work on its modelling, awaiting to have a good representation of the features (c) and (d) to subtract. A possible explanation is particle induced Bremsstrahlung, and in fact Xe cross section increases by a factor 3 going across the L edge. However it is difficult to explain a positive slope (i.e. intensity higher at high energies). We note however that this slope depends on the BL cuts in use. Our results are referred to the standard BL thresholds [2]. Using narrower BL limits, one can obtain a flatter, or even decreasing, distribution.

## 4. COSMIC BACKGROUND

We hope to be able in the future to produce a model representative of the instrumental background (vs energy and position) to be subtracted from the blank field spectra in order to produce net cosmic background spectra (component (f)). At the moment we can only resort to a subtraction of the observed dark Earth spectra, whose error bars (see Fig. 6) are however dominated by the statistics of the latter, because of the shorter exposure time (see Table 1).

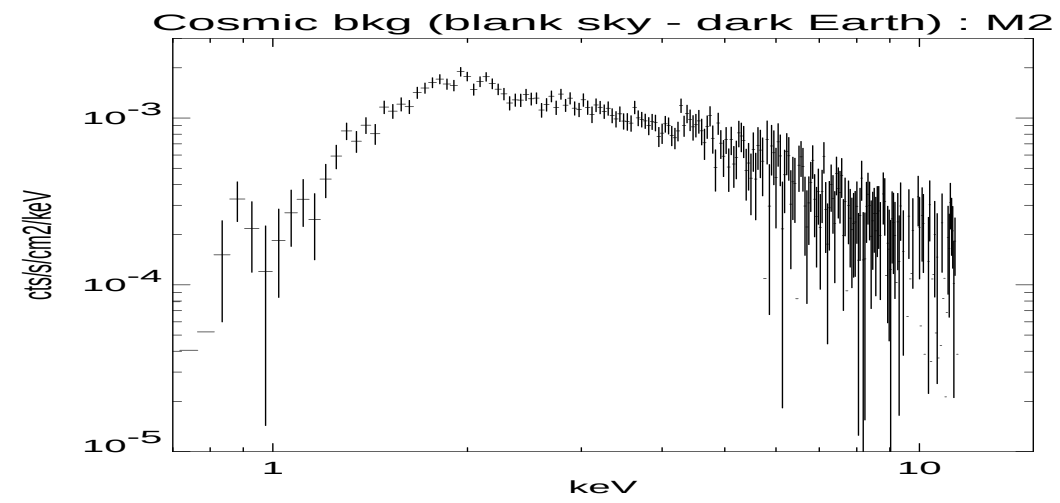

Figure 6. The net spectrum of the cosmic background (after subtraction of the dark Earth insturmental contribution) for MECS unit M2, normalized to the area of the entire FOV.

As a possible interim solution in the future we could perhaps average dark Earth spectra with bright Earth spectra (cleaning out component (e) features).

The resulting spectra (see Fig. 6) show a spatial modulation (see Fig. 7) resemblant of the vignetting of the optics [2,5].

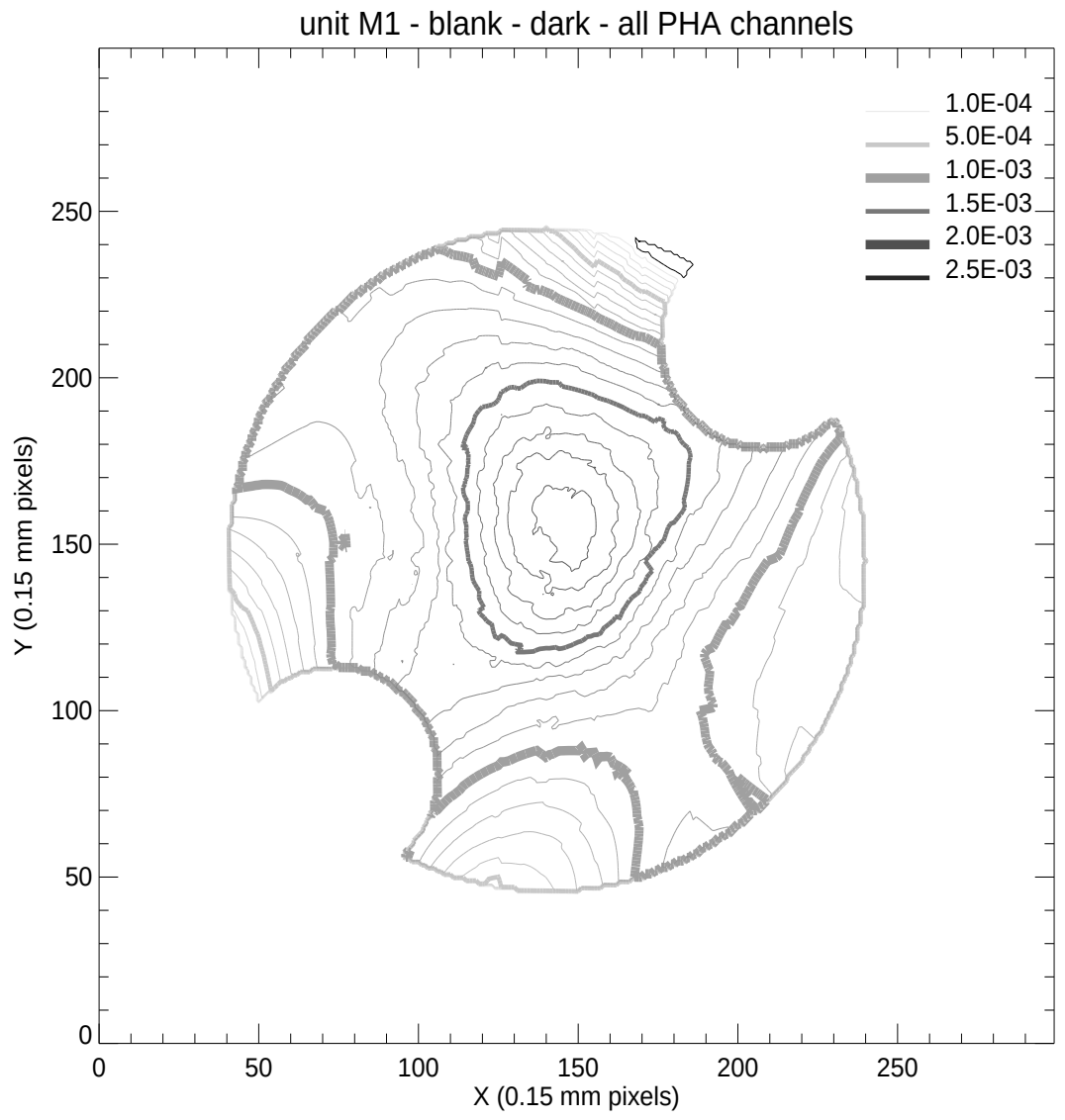

Figure 7. Example of spatial modulation of the net cosmic background : count rate in the total band for detector M1.

## 4.1. The response matrix

In order to perform a spectral fitting of the net cosmic background spectra, one shall fully understand all position-dependent effects entering in the response matrix, customarily decomposed into an RMF (Response Matrix File) and an ARF (Ancillary Response file) :

$$Q(E,PHA)=RMF(E,PHA)ARF(E)$$

In turn the ARF is the product of a number of components [2]:

(i) on axis optics effective area $A_{onaxis}(E)$
(ii) optics vignetting function $V(E,x,y)$
(iii) Panter-to-infinity correction $I_{\infty}(E,x,y)$
(iv) transmission of plasma grid $\tau_g$
(v) transmission of the UV/ion filter $\tau_f(E)$
(vi) transmission of the Be window $\tau_w(E,x,y)$
(vii) detector quantum efficiency $\varepsilon(E)$
(viii) burst length correction $B(E)$
(ix) optics+detector PSF fraction enclosed in extraction radius $f(E,r_{ext})$

We have neglected component (ix), which in principle has a loose positional dependency through the shape of the PSF, since in the case of the background we are dealing with an uniform illumination and there is no such thing as an extraction radius.

The only other components which are positional dependents are (ii) and (iii), which can be combined in a single radial function, and (vi). $\tau_w(E,x,y)$ can be expressed as :

$$P_{la}(E,x,y)e^{-\sigma(E)t_{la}} + (1-P_{la}(E,x,y))e^{-\sigma(E)t_{st}}$$

where $t_{la}$ and $t_{st}$ are the thicknesses of the Be layer and of the supporting strongback [2], while the only position-dependent parts are $P_{la}$ (the coverage fraction of the Be layer, i.e. the convolution of a 1/0 mask representing the entrance window with the optics PSF) and its complement $1-P_{la}$ (the coverage fraction of the strongback). Note that this latter component has not a radial symmetry.

We have developed a modified response matrix generation program, which, given a generic region, computes point by point the product of components (ii),(iii) and (vi) and averages them on the region, then computes once the product of the RMF times the remaining ARF components times the above average.

We have used matrices generated with the above program in the 35 boxes to simulate individual box spectra and generate from them an intensity map (where the presence of the strongback is quite washed out and remains only as a perturbation of the overall symmetry) which is in good qualitative and quantitative agreement with the map derived from actual measurements.

## 4.2. Fitting results

We hoped to be able to apply the position-dependent matrix generator for fitting of individual box spectra, in order to verify (or improve) the calibration of the vignetting and of the Be window transmission. However the statistics of the individual box spectra (particularly the outermost ones) does not allow a very good quality of the results. We have therefore repeated the accumulation of blank field and dark Earth spectra in selected regions, as indicated in Table 2.

Table 2
Regions used for background accumulation

| Description | radius (arcmin) | solid angle (sr) |
|---|---|---|
| a) circle circumscribed to inner box D4 | 5.8 | $9\times10^{-6}$ |
| b) circle used in [6] | 8.4 | $1.9\times10^{-5}$ |
| c) circle inside regions of avoidance of calibration sources | 16.7 | $7.4\times10^{-5}$ |
| d) outer ring ("e"-complement of "c") | 20[a] | $10^{-4}$ |
| e) entire FOV | 26[a] | $1.8\times10^{-4}$ |

[a] equivalent radius (non circular region)

The resulting spectra (rebinned to achieve at least a 5σ signal in each bin) are shown in Fig. 8. We attempted to fit them (using the appropriate position-dependent matrix) with a power law of the form

$$dN/dE d\Omega = N E^{-\alpha} \ \mathrm{ph/cm^2/s/keV/sr}$$

Considering that we used all blank fields with no selection in galactic coordinates, and the fact that the MECS are not particularly sensitive to the hydrogen column density, we fixed $N_H=5\times10^{20}$ cm$^{-2}$.

We encountered the following problems while fitting.

The results of the innermost region are in good agreement with the current model (N= 11 ph/cm$^2$/s/keV/sr ; photon index 1.4, see [7,8]), but the confidence contours are quite large (and elongated since the fitting parameters are correlated, because the normalization is computed at 1 keV, just outside our fitting range.). As apparent in Fig. 9, if we use the outer regions the statistics improves, and the confidence contour narrows, but the normalization increases to 15 ph/cm$^2$/s/keV/sr.

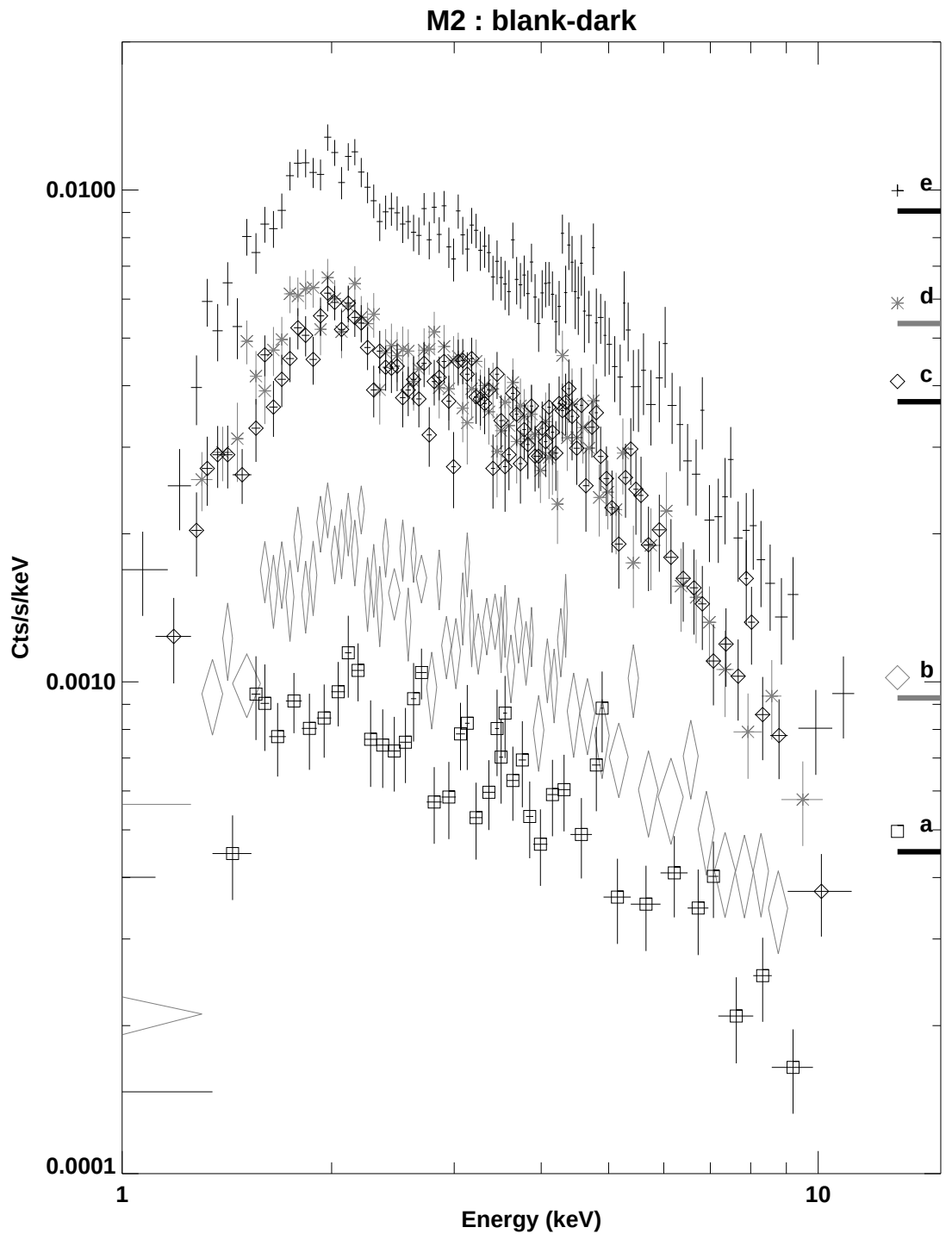

Figure 8. Net spectra of the cosmic background in the regions listed in Table 2 (symbols indicated on the right together with fiducial marks indicating solid angles on arbitrary scale).

This might point out the need of improvements to the knowledge of the calibrations (vignetting, PSF) or a misappreciation of effects of stray X-rays from outside the field of view (though small can be the effect, see [5]).

Another problem is that we obtain the expected photon index if we fit only in the 3-10 keV range. If we extend the fitting range down to 2 keV we obtain bad fits (particularly with MECS units M2 and M3 which have a thinner Lexan filter with respect to the Kapton one mounted on unit M1), and conversely if we extrapolate the α=1.4 fit obtained in 3-10 keV range to lower energies, the latter data points remain above the fit.

We do not dare so far to ascribe such preliminary results to a softer component or excess (somewhat unexpected above 1 keV, see [7,8]). We considered that we used all blank field data together, inclusive of three long secondary pointings taken during 21-30 August 1996 towards the Draco region (in whose neighbourhood a true excess may be present). We repeated our spectra accumulations separately for the Draco pointings (517 ks) and for all other pointings (560 ks), and note that the resulting spectra are remarkably similar except *below* 2 keV. This could be due just to the lower mean $N_H$ towards those directions ($2\times10^{20}$ $cm^{-2}$).

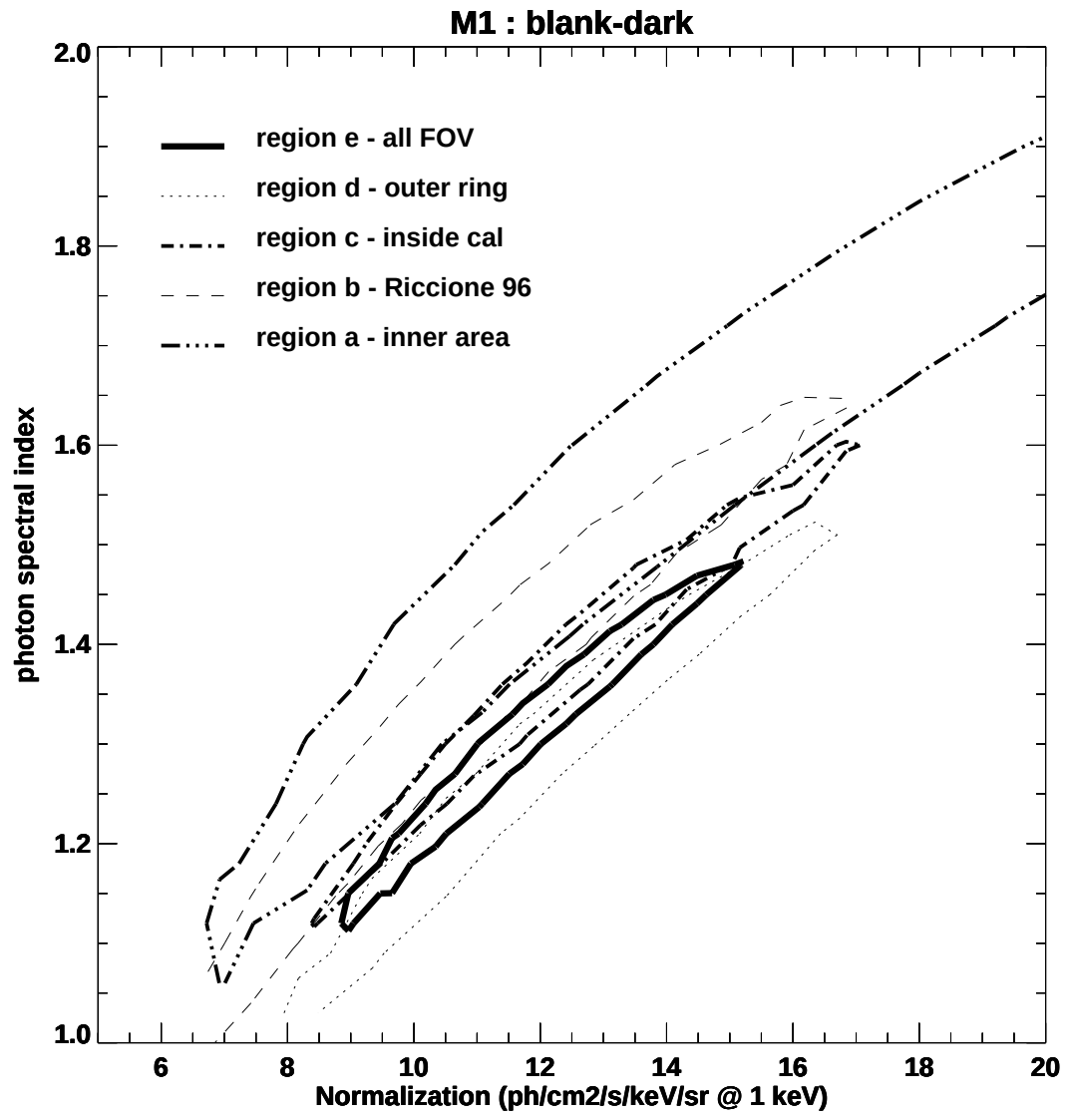

Figure 9 : 99% confidence contours of the best fits to M1 spectra (accumulated in the regions listed in Table 2).

## 5. CONCLUSIONS

The present preliminary work proves that the MECS is potentially capable of giving a good spectral fit of the cosmic X-ray background. Further work is necessary (and worthwhile) to improve the statistics of the subtraction of the instrumental background. Inclusion of data of the entire field of view also potentially improves the quality of the fit (as shown by Fig. 9). We need however a better understanding of the spatially dependent components of the response matrix, and possibly a more careful selection of which blank fields to combine according to their galactic coordinates.

## ACKNOWLEDGMENTS

We thank R.C.Butler and L.Piro, *BeppoSAX* flight director and mission scientist, and the staff of Nuova Telespazio for their support to the SVP activities.. We wish to recall here all members of the MECS team in the scientific institutes, who participated to its development and calibration, and in particular G.Boella, S.Re, G.Conti, G.La Rosa and B.Sacco.

The present research has been accomplished despite the actions by the Italian governmental bureaucracy (and in particular by ARAN) to degrade the work conditions of research staff of Public Research Organizations (PRO), like CNR, and of their attempt to introduce a radical disruption of the unity of research between PRO and Universities.